# Risk Assessment of Multi-timescale Cascading Outages based on Markovian Tree Search

Rui Yao, Shaowei Huang, Kai Sun, Feng Liu, Xuemin Zhang, Shengwei Mei, Wei Wei, Lijie Ding

*Abstract*— In the risk assessment of cascading outages, the rationality of simulation and efficiency of computation are both of great significance. To overcome the drawback of sampling-based methods that huge computation resources are required and the shortcoming of initial contingency selection practices that the dependencies in sequences of outages are omitted, this paper proposes a novel risk assessment approach by searching on Markovian Tree. The Markovian tree model is reformulated from the quasi-dynamic multi-timescale simulation model proposed recently to ensure reasonable modeling and simulation of cascading outages. Then a tree search scheme is established to avoid duplicated simulations on same cascade paths, significantly saving computation time. To accelerate the convergence of risk assessment, a risk estimation index is proposed to guide the search for states with major contributions to the risk, and the risk assessment is realized based on the risk estimation index with a forward tree search and backward update algorithm. The effectiveness of the proposed method is illustrated on a 4-node power system, and its convergence profile as well as efficiency is demonstrated on the RTS-96 test system.

## NOMENCLATURE

*Symbols*

| | |
|---|---|
| $R$ | Cascading outage risk |
| $C_0$ | Immediate cost after initial outage |
| $i_k$ | Index of element $k$ |
| $i_{k_1} i_{k_2} \cdots i_{k_r}$ | Cascading outage denoted by mid-timescale element outage sequence |
| $(i_{k_1} i_{k_2} \cdots i_{k_r})$ | System state after outage sequence $i_{k_1} i_{k_2} \cdots i_{k_r}$ |
| $\Pr(i_{k_1} \cdots i_{k_r})$ | Probability of cascading outage $i_{k_1} i_{k_2} \cdots i_{k_r}$ |
| $\Pr(i_{k_{r+1}} \mid i_{k_1} \cdots i_{k_r})$ | Conditional probability of element $i_{k_{r+1}}$ outage after outage sequence $i_{k_1} i_{k_2} \cdots i_{k_r}$ |
| $C(i_{k_1} i_{k_2} \cdots i_{k_r})$ | Cost at the state after outage sequence $i_{k_1} i_{k_2} \cdots i_{k_r}$ |
| $\tau_D$ | Length of mid-timescale interval |
| $\lambda_i$ | Outage rate of element $i$ |
| $\Pr_i^{MC}$ | Probability of element $i$ outage in sampling-based model |
| $\Pr_i^{MT}$ | Probability of element $i$ outage in Markovian tree model |
| $\Delta t_{Delay}$ | Re-dispatch delay |
| $Y$ | Admittance matrix |
| $Z$ | Inverse of admittance matrix |
| $Y_{uv}, Z_{uv}$ | Element at the position of $(u,v)$ in matrix |
| $E$ | Set of working elements |
| $V$ | Set of buses |
| $E_{i_{k_1} \cdots i_{k_r}}$ | Set of working elements after outages $i_{k_1} \cdots i_{k_r}$ |
| $\{u,v\}$ | Branch with node $u$ and $v$ as terminals |
| $F_{uv}$ | Power flow on branch $\{u,v\}$ |

*Abbreviations*

| | |
|---|---|
| UVLS | Under-voltage load shedding |
| PTDF | Power transfer distribution factor |
| TLR | Transmission loading relief |

This work was supported by the Foundation for Innovative Research Groups of the National Natural Science Foundation of China (51321005), the National Science Foundation of China (51377091) and the Electric Power Research Institute of Sichuan Province.

R. Yao, S. Huang, F. Liu (corresponding author), X. Zhang and S. Mei are with the State Key Laboratory of Power Systems, Department of Electrical Engineering, Tsinghua University, Beijing 100084, China (emails: yaorui.thu@gmail.com, lfeng@tsinghua.edu.cn).

K. Sun is with the Department of EECS, University of Tennessee, Knoxville, TN 37996, USA (e-mail: kaisun@utk.edu).

W. Wei and L. Ding are with the State Grid Sichuan Electric Power Company, Chengdu, Sichuan 610041, China.

## I. INTRODUCTION

R ISK assessment of cascading outages in power systems is an important topic since cascading outages are big threats to electricity supply as well as to the society[1-3]. One of the most challenging problems of risk assessment is the limitation of calculation speed brought by the huge number of possible cascade paths and the lack of computation resources. One commonly-utilized approach of risk assessment is the Monte-Carlo method, which repeatedly creates samples of events based on their real probabilities until risk converges. However, Monte-Carlo method requires large numbers of samples to converge, especially for rare events [4]. Since the convergence of sampling-based methods relies on the variance of sampling, various variance reduction methods are proposed, e.g. importance sampling [5], cross-entropy [6], stratification [7], etc. These methods can accelerate computation by several times compared to Monte-Carlo method. However, the improvements are limited and risk assessment still requires huge computation.

The selection of serious contingency patterns is another approach for risk assessment. Various contingency or state combination selection techniques have been proposed [8-10]. However, these methods treat outages as independent of each other, which neglects the fundamental "cascading" nature of cascading outages. Reasonable risk assessment requires updating the state as well as probability of element outages during simulation, which makes the techniques utilized in contingency selection methods ineffective.

Correctly capturing the dependency within sequences of outages is a prerequisite of reasonable and practical risk assessment, so the simulation of cascading outages is inevitable in risk assessment. An effective way of efficiency enhancement is to reduce the invocation of simulation. In sampling-based methods, lots of time is wasted in duplicated sampling of same cascade paths. Actually upon knowing the conditional probabilities on each level of cascading outages (which is feasible in most existing cascading outage models), the efficiency of risk assessment can be significantly improved by simulating the cascading outage path and directly estimating risk indices using the probability, which avoids duplicated simulation of same paths. All possible cascading outage paths can be formulated as a tree structure. As cascading outages can be regarded as Markovian processes [11, 12], risk assessment can be realized as searching on a Markovian tree [13].

Ref. [11] proposes a Markovian tree model of cascading line outages. The model considers the effect of random load fluctuation and incorporates time into simulation, which enhances practicality. However, the distinct mechanisms of line outage directly caused by triggering protection and by slow outage process [14] caused by tree contact are not distinguished. And process as slow change of load level and generation re-dispatch are not considered.

Risk assessment should be based on reasonable modeling and simulation. Various models have been proposed, including topology-based models [15], high-level statistical models [16-18], power-flow based models [19-21], models treating cascading outages in different stages [22], and models mainly considering severe initial outage combinations [23], etc. To reflect the essence of cascading outages as a multi-timescale complex process which is essential for reasonable simulation and risk assessment, our previous work [24] proposes a multi-timescale quasi-dynamic simulation model. In the model, the events in cascading outages are categorized into three timescales, and the interactions among timescales as well as the representation of time elapse are realized within the quasi-dynamic framework, which enhances the practicality.

Based on the model in [24], this paper proposes a risk assessment method of multi-timescale cascading outages based on Markovian tree search. The multi-timescale modeling and quasi-dynamic simulation of cascading outage paths are reformulated as a Markovian tree, and then the risk assessment is realized by searching on Markovian tree. Since duplicated simulation of same paths is avoided, the efficiency is expected to be significantly enhanced. According to the analysis of the causes of the risk, a searching strategy based on a risk estimation index is proposed to speed up convergence of risk, which further accelerates risk assessment.

The major contributions of this paper are listed as follows:

• A multi-timescale cascading outage simulation model based on a structure of Markovian tree is established. This model is equivalently transformed from our previous work [24] of a multi-timescale simulation model with representation of time, which is more reasonable than existing models with tree structure. Also the detailed realization of the proposed model has non-trivial difference from [24], such as the tree structure and the definition of outage probability. The proposed model maintains the merits of model [24], and bridges the reasonable modeling with efficient risk assessment based on tree search.

• Based on the Markovian tree modeling and simulation of cascading outages, a non-duplicated risk assessment methodology using Markovian tree search is proposed. The novel risk assessment method avoids duplicated simulation of cascading paths, significantly saving computation resources. The convergence criteria of risk assessment are also proposed.

• A novel forward-backward Markovian tree search scheme based on a risk estimation index is proposed, further accelerating the convergence of risk assessment. The risk estimation index is established by analyzing the causes of cascading outage risk and estimating the costs through theoretical derivation, which effectively facilitates the selection of high-risk cascading outage paths, significantly accelerating the convergence of risk assessment. Moreover, the risk estimation index is updated reversely considering the influence of simulated states, forming an efficient forward-backward Markovian tree search algorithm.

The rest of this paper is organized as follows. Section II reformulates the quasi-dynamic multi-timescale simulation as a Markovian Tree. Section III proposes the idea of risk assessment with Markovian Tree search. Section IV proposes a risk estimation index for guiding of the search. And then a forward-search-backward-update scheme of risk assessment based on the risk estimation index is established. Section V presents test case studies of the risk assessment method on a 4-node system and RTS-96 test system. Section VI draws the conclusion.

## II. CASCADING OUTAGE SIMULATION WITH MARKOVIAN TREE

### A. Brief Retrospect of Multi-Timescale Simulation Model

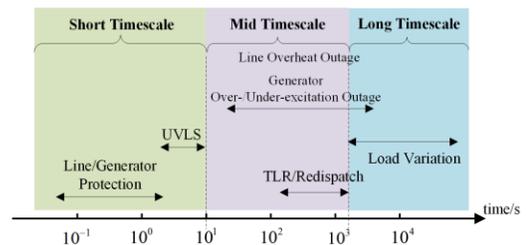

Fig. 1. Timescales of dynamics in cascading outages [24].

To represent the time elapse in simulation, ref. [24] categorizes events in cascading outages by timescales, as shown in Fig. 1. Then a quasi-dynamic simulation framework is utilized, which inserts the simulation of shorter timescales between adjacent longer-timescale events (Fig. 2). Thus multi-timescale simulation is realized and approximate time elapse is provided, which reasonably reflects the actual characteristics of cascading outages. Since reasonable modeling and simulation of cascading outages is a prerequisite of practical risk assessment, next we will show how to reformulate the quasi-dynamic multi-timescale model as Markovian tree as a foundation of the novel risk assessment method.

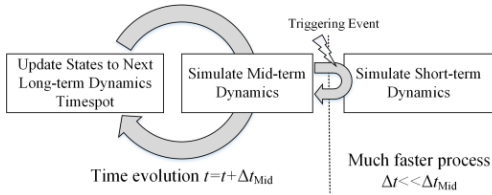

Fig. 2. Quasi-dynamic multi-timescale simulation framework [24].

### B. Markovian-Tree Simulation of Cascading Outages

#### 1) Markovian tree modeling of mid-timescale random outages

In the model proposed in [24] (named as "quasi-dynamic model" in the rest of the paper), the mid-timescale random outages have uncertainties, causing distinct cascading outage sequences. If we merge same states of all the possible cascading outage sequences from the beginning state, a tree structure will be formulated, as shown in Fig. 3. Assume cascading outages are Markovian [11, 25], the tree is then a Markovian tree. Each node on Markovian tree represents a state, and each branch on Markovian tree represents a mid-timescale random outage. Similar to the quasi-dynamic model, every mid-timescale transition corresponds to a time elapse $\tau_D$, and the reasonable values of $\tau_D$ will be discussed afterwards with test cases.

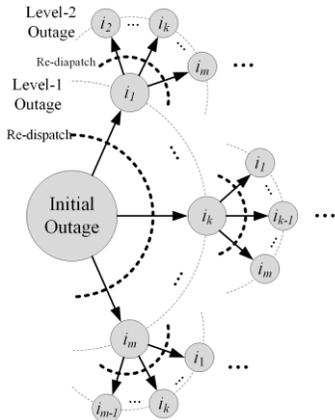

Fig. 3. Markovian tree representation of cascading outage paths

The labelling of states on a Markovian tree is determined as follows. The beginning state is level-0 node, and subsequent states are labelled sequentially as level-1 nodes, level-2 nodes, etc. Note that not every state transition corresponds to an outage, it is possible that no outage occurs during $\tau_D$. If we label elements with positive integers, then a state is coded as the sequence of labels of failed elements from the initial outage up to the current level (no-outage labelled as 0), i.e. $(i_{k_1} i_{k_2} \cdots i_{k_n})$. In this paper only branch outage events are considered, but the modeling and simulation of generator outages are very similar.

The costs of cascading outages come from the dispatch control, load shed by power balancing when system separates, or load shed by emergency control measures on each level of the Markovian tree. The selection of the definition of the cost also has flexibility: the cost can be the load or the energy loss at each level of outage, or the economic loss caused by the outages. In this paper, the cascading outage simulation is based on DC power flow, and the cost of an outage $C$ on Markovian tree is the sum of *load loss from re-dispatch* $C_R$ and *load loss from network balancing* $C_B$. Every state $(i_{k_1} i_{k_2} \cdots i_{k_n})$ corresponds to a non-negative cost denoted as $C(i_{k_1} i_{k_2} \cdots i_{k_n})$.

$$C(i_{k_1} i_{k_2} \cdots i_{k_n}) = C_R(i_{k_1} i_{k_2} \cdots i_{k_n}) + C_B(i_{k_1} i_{k_2} \cdots i_{k_n}) \quad (1)$$

The probability of any event $i_{k_{n+1}}$ (corresponding to state $(i_{k_1} i_{k_2} \cdots i_{k_{n+1}})$) depends on its previous state $(i_{k_1} i_{k_2} \cdots i_{k_n})$, so the conditional probability of event $i_{k_{n+1}}$ is denoted as $\Pr(i_{k_{n+1}} | i_{k_1} i_{k_2} \cdots i_{k_n})$. With these terms, take the risk assessment of expected load loss as example, the risk is expressed as

$$
\begin{aligned}
R = C_0 &+ \sum_{k_1} \Pr(i_{k_1}) C(i_{k_1}) + \sum_{k_1} \Pr(i_{k_1}) \sum_{k_2} \Pr(i_{k_2} | i_{k_1}) C(i_{k_1} i_{k_2}) \\
&+ \sum_{k_1} \Pr(i_{k_1}) \sum_{k_2} \Pr(i_{k_2} | i_{k_1}) \sum_{k_3} \Pr(i_{k_3} | i_{k_1} i_{k_2}) C(i_{k_1} i_{k_2} i_{k_3}) + \cdots
\end{aligned}
\quad (2)
$$

Different from the quasi-dynamic model which is able to sample more than one outage in each interval, here similar to [11, 25, 26], each interval $\tau_D$ allows at most one element outage. To ensure the equivalency between the quasi-dynamic model and the Markovian tree model, the mid-timescale interval $\tau_D$ in the Markovian tree is set as $1/N_\tau$ of that in the quasi-dynamic model, thus the Markovian tree model is equivalent to the case of sampling up to $N_\tau$ outages during the same period. It is found through tests that generally $N_\tau = 3 \sim 5$ can satisfy practical needs. Moreover, the Markovian tree model considers the sequence of outages, which is even closer to system reality than the quasi-dynamic model.

The requirement of single outages on the Markovian tree renders different definitions of probability. In the quasi-dynamic model, outage events are sampled independently, while in the Markovian tree model, the outage probability is defined as "the probability that the outage is the first to occur". Assume that the occurrence of outages follow Poisson process, where the outage rate of element $i$ is $\lambda_i$. Then in the quasi-dynamic model, the outages are sampled independently with probabilities in interval $\tau_D$:

$$\Pr_i^{Sample} = 1 - e^{-\lambda_i \tau_D} \quad (3)$$

while in the Markovian tree model, the outage probability of each element is

$$\Pr_i^{MT} = \frac{\lambda_i}{\sum_j \lambda_j} \left( 1 - e^{-\sum_j \lambda_j \tau_D} \right) \quad (4)$$

where "MT" is the abbreviation of Markovian tree. In (4) the outage probability of element $i$ is not only dependent on $\lambda_i$, but also on outage rates of other elements. The probability that there is no outage in interval $\tau_D$ is

$$\Pr_0^{MT} = e^{-\sum_j \lambda_j \tau_D} \quad (5)$$

#### 2) Simulation of Re-dispatch

Re-dispatch is categorized as a mid-timescale process. When overload occurs, dispatchers adjust generators or dump loads to relieve the overload. In conventional models [19, 20, 22], re-dispatch is modeled as an optimization problem and the optimal solution is instantly applied as the new state.

However, re-dispatch takes time [27]: when overloading occurs, the system needs time to acquire data, analyze system conditions and reflect the data to operators; the operators also need time to judge and make decisions before taking actions. Moreover, due to the generation ramping speed constraints, it also takes time from the beginning of actions till the fulfillment of re-dispatch objectives. Therefore, the re-dispatch is a process with a time delay $\Delta t_{Delay}$ and ramping.

As shown in Fig. 4, when an overloading event occurs at $t_0$, the re-dispatch action for the event is not immediately started. During interval $t_0 < t < t_0 + \Delta t_{Delay}$, the system may be taking actions dealing with previous events or there is no action at all. Considering the time-delay nature of re-dispatch, a queue of re-dispatch commands is prepared in simulation. As an event occurs, add the corresponding command to the queue and wait until the action is due. The latest command meeting the beginning time is offered from the queue and starts executing. The command in action is kept until re-dispatch is finished or it is replaced by a new command.

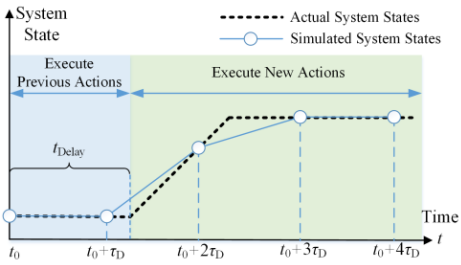

Fig. 4. Illustration of Re-dispatch Simulation

*3) Simulation of Short-timescale Processes*

Ref. [24] mentions that short-timescale processes mainly refer to outages directly triggered by protection relays and actions of emergency control. These processes usually finish in several seconds and are much shorter than processes in other timescales, and these processes follow strict preset logics.

In simulation, when system states change, check whether short-timescale processes occur, if so then first simulates them. As illustrated in Fig. 5, event $i_{k_m}$ triggers a short-timescale event denoted as $i_{l_1|k_m}$, and consequently triggers event $i_{l_2|k_m}$, then afterwards the short-timescale process ends. Since this process is very short compared with the Markovian tree structure of mid-timescale processes, the short-timescale process can be modeled as an equivalent node.

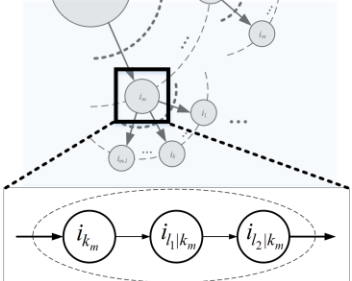

Fig. 5. Illustration of Short-timescale Processes in Markovian tree

In the simulation of short-timescale processes, there may be load losses caused by island balancing when system separates or by emergency load shedding. It should be noted that since these losses are not caused by market-based measures, the unit economic cost of these losses are usually much higher than electricity market prices. Therefore when estimating the expected economic loss, the unit economic cost is $\mu_E$ (e.g 100[28]) times of dispatch operations.

It should also be noted that this model can deal with more diverse events, such as bus outages and instability events. The simulation of bus outages is similar to that of short-timescale branch/generator outages. The instability events usually need time-domain simulation [24, 29], which can also be incorporated into this simulation model.

*4) Simulation of long-timescale processes*

A long-timescale process corresponds to the variation of load level. Since searching in depth on the Markovian tree also means elapse of time, the simulation of load variation can be realized by updating system loads according to the load curve.

### C. An illustrative example of Markovian tree

To better illustrate the mechanism of cascading outage simulation with Markovian tree, here an example is provided. As Fig. 6 shows, the cascading outages start with initial outage(s) at time $t = 0$, then the initial outage(s) might trigger short timescale events and re-dispatch operation. Since short-timescale events are much faster than the re-dispatch and mid-timescale outages, the short-timescale events should be the first to be simulated. After outages and short-timescale events, the re-dispatch operation and mid-timescale outage in an interval $\tau_D$ are simulated.

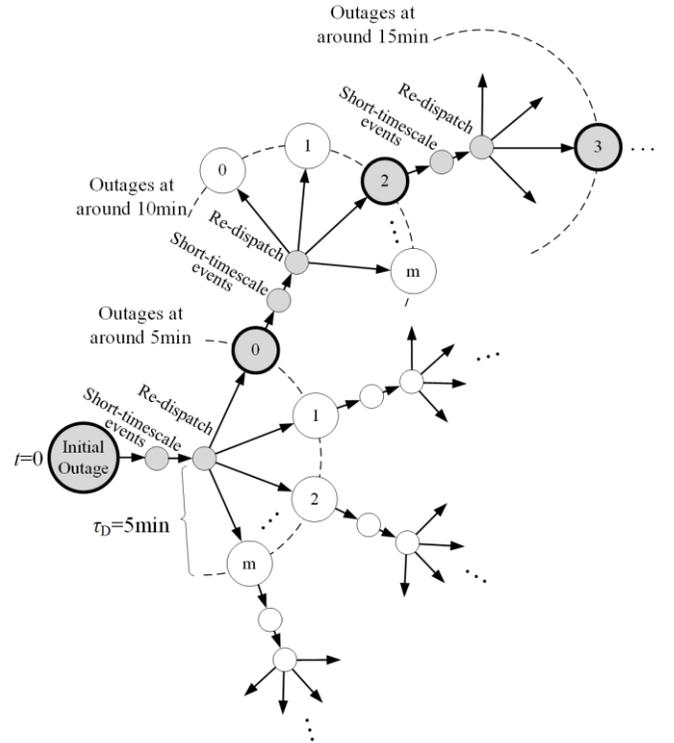

Fig. 6. An illustrative example of Markovian tree

Because any element may fail in an interval $\tau_D$, so there are various possible directions of cascading outage development, corresponding to the forked structure in every interval shown in Fig. 6. The direction of cascading outage development is denoted with the index of the failed element. Note that there might be no outage during an interval, and in this case the path is indexed as 0. So during the interval $\tau_D$ after any state, there is a forked structure of possible cascading outage directions. Therefore, all the possible cascading outage paths can be collected as a tree structure as Fig. 6 demonstrates.

In each cascading outage simulation with Markovian tree, a complete cascading outage path is simulated from the initial outages till the terminal of the path. On the Markovian tree structure, the simulated path is a linked list of nodes on the tree starting from the root node (initial outages) to the terminal. In Fig. 6, the filled nodes constitute a cascading outage path. The path means that at around 10 minutes after initial outages, the element 2 fails, and at around 15 minutes the element 3 fails.

It should be noted that due to the dependency in a cascading outage sequence, each post-outage state should be updated rather than regarding the outages as independent combinations. And since there are possibly re-dispatch operations, emergency control measures and dynamic processes after each outage, the outcomes of different orders of outages from the same outage combination are probably different, and the risks of these different outage sequences are probably different. So in this paper the outage sequences in different orders are treated as different and are simulated respectively.

It is also remarkable that the proposed Markovian tree model maintains the merits of model [24], yet as the previous sections demonstrate, the realization of the two models still has substantial difference and the transformation from model [24] to the Markovian tree is not trivial work. The following sections will show that the efficient risk assessment method will be realized based on the Markovian tree structure. So the Markovian tree model plays as a bridge between reasonable simulation of cascading outages and realization of efficient risk assessment.

## III. ENHANCING EFFICIENCY USING TREE SEARCH

### A. The methodology of risk assessment with tree search

Following the details proposed in Section II, simulation of cascading outages can be realized on the Markovian tree as equivalent to the model in [24], and risk assessment can be carried out by sampling on the Markovian tree. Yet sampling duplicates simulation of same cascade paths. To enhance efficiency, the simulated cascade paths can be recorded and avoided in further simulation. Thus risk assessment becomes searching on the Markovian tree.

The risk (2) can be regarded as the sum of risk terms of all the single states on the Markovian tree. Therefore, the risk assessment based on the Markovian tree can be regarded as simulating new states on the Markovian tree and adding new corresponding terms onto (2). Since the terms in (2) are non-negative, the risk is expected to keep increasing until reaching a value $R$ which is the cascading outage risk of the system.

Fig. 7 illustrates the mechanism of risk assessment with Markovian tree search. The nodes with bold borderline constitute a partial cascading outage path. The nodes with grey color denote states that have been simulated in previous searches. In current search, these states are directly retrieved from the memory. The nodes with white color are states that have not been reached. The events of these states are simulated based on section II, the risk terms corresponding to these states are added to (1), and then the states are stored in the memory.

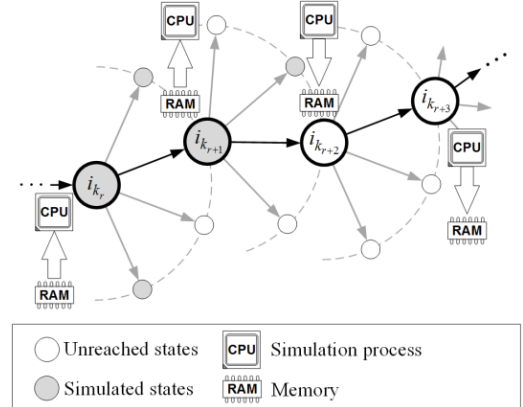

Fig. 7. Risk assessment with Markovian tree search

The risk assessment using tree search has similarities with the splitting method [30, 31]. The splitting method enhances the efficiency of cascading outage simulation by splitting the tree structure of cascade events into stages, and initiating the sampling of a stage from the desired sampled ending states of the previous stage which is stored in the memory. In essence, the splitting method and the Markovian tree search method both make use of the data generated in the previous computation, and manage to achieve time efficiency enhancement with some sacrifice on memory space. In the analysis of cascading outages, the computation time is more critical than memory resources, so exploiting memory to enhance time efficiency is reasonable.

However, the realization and the assumption of these two methods are different. The splitting method is still based on sampling the random outages, and the Markovian tree search method assumes that the probability of any outage can be obtained and the risk is directly estimated using the probability. In Markovian tree search method, since the risk term is estimated with only one simulation, so each state can be only simulated once. Whereas for sampling-based models, repeated simulations of same cascade paths is inevitable. Therefore, the Markovian tree search method is more efficient than the sampling-based models. It should be noted that the Markovian tree search method cannot deal with the case that the random outage is a black box and the probability cannot be explicitly obtained. The probabilistic modeling in cascading outages will be more extensively discussed in III.B.

While comparing the memory usage of splitting method and the Markovian tree method, the splitting method saves more memory space than the current realization of the Markovian tree method, since the splitting method is able to discard the results of previous outage sequences and store the end states only. However, as the test case in Section V.B will show, the memory usage of the Markovian tree search method can be satisfied with common computation platforms. Moreover, such difference is on the realization level and is caused by the different purposes of these two methods. From the perspective of the event tree, the splitting method in [31] mainly focuses on the simulation and estimation of rare and extreme events, which requires to search in-depth into the event tree. So the partial event tree formed by splitting method can be very deep but is relatively thin. Whereas in this paper, the Markovian tree method is proposed to estimate the overall risk of cascading outages, the event tree should first be wide enough to cover as fully as possible the lower-level events, because these events have higher probabilities even though the costs of these event are usually small, and thus the risks of these lower-level events cannot be omitted. And the in-depth searching to high risk cascading outage sequences depends on the strategy to be introduced in Section IV. Both of the two methods have potential for alternative purposes. For example, in the splitting method, more times of sampling can be arranged in the lower levels of cascading outages to adapt to the need for risk assessment. And the Markovian tree search can also be realized in segments to focus on the rarer and more extreme events.

### B. Discussion on Probabilities in Risk assessment

The probabilities of outages are derived according to the model [24] that considers the time elapse of heat accumulation and some environmental factors. Yet it should be noted that in practice, estimating the probabilities of cascading outages is a difficult problem that has not been well solved. The difficulties of estimating the probability of cascading outages mainly lie in the following factors:

1) There are various kinds of causes to outages in a whole process of cascading outages. A power system consists of many kinds of elements, and any element may outage during the process of cascading outages due to various causes [32]. A line may outage due to sagging (2003 US-Canada blackout), annealing, lightning (2009 Brazil-Paraguay blackout), mis-operation (2011 Arizona-California outages), improper protection setting (2006 UCTE disturbance), power system oscillation (1996 WSCC outages), ice and snow (2008 South China blackout), geomagnetic storm (1989 Hydro-Quebec blackout), etc. A generator may fail due to over-current, over-/under-excitation, over-/under-frequency, etc. Also there are other kinds of elements or components that may outage, such as underground cables, transformers (2005 Moscow blackout), HVDC, control center (2003 US-Canada blackout) and communication infrastructure (1988 Hydro-Quebec blackout), etc. The power system itself also demonstrates complex dynamics in multiple timescales during cascading outage. Therefore, the cascading outage is a process involving complex and dependent system behavior, and a variety of outages of different kinds of elements caused by system dynamics and/or external factors. To comprehensively study all kinds of events in cascading outages, it is necessary acquire adequate amount of data and logs in operation. However, currently the power systems still face lack of data, which hinders credible modeling and verification of possible kinds of element outages.

2) The modeling of each kind of outage event is difficult due to various influencing factors. Currently, providing probabilistic model of each kind of outage mechanism is difficult because of (a) the difficulty in establishing the probabilistic model considering all the important influencing factors and (b) the lack of sources to collect data of those factors in practice. Take the outage of overhead line caused by sagging and tree contact as example, the IEEE Standard 738 [33] points out that there are various factors that influences the steady-state sag of overhead lines, including the current on conductor, the type of conductor, ambient temperature, wind speed, wind direction, season, the time of day, sunshine illumination, etc. Moreover, the outage caused by tree contact also depends on the height of vegetation under the overhead line. We can see that not only electric-side variables but also many environmental factors influence the line outage event, while till now the line outage models in the existing literatures [19, 24, 26, 34, 35] are intuitive but not accurate enough, and there is not yet credible probabilistic model of such outage event with comprehensive consideration of the influencing factors [36]. Moreover, in practice, it is difficult to monitor or predict all these environmental factors accurately, making the model hard to utilize in application.

From above, it can be concluded that the estimation of outage probability requires more accurate modeling of the physical process of various kinds of outages, which asks for enhanced study of element reliability, modeling of environmental factors as well as verification of models with field tests and observations during operation. Also, concentration should be placed on enhancing the situational awareness of power systems to enable the utilization of more accurate models, especially for the monitoring and management of environmental data.

### C. Convergence criteria of risk assessment

In the risk assessment with Markovian tree search, the theoretical value of risk is obtained when all the possible cascading outage paths are exhausted and simulated. if there are $N$ elements at the initial state, and $K_D$ levels on the Markovian tree, then the number of all possible cascade paths on Markovian tree is

$$N_T = \sum_{i=0}^{K_D} \frac{N!K_D!}{(N-K_D+i)!(K_D-i)!i!} \qquad (6)$$

From (6) we can see that the number of possible cascade paths is huge for common-sized systems so that it is practically impossible to exhaust all the paths. Therefore, in practice we can only simulate a portion of the paths. In risk assessment, as cascade paths are simulated, the value of risk keeps growing and gradually approaches the theoretical value, and the total probability of simulated cascade paths also approaches 1. Thus the criteria for the convergence of risk assessment are proposed as satisfying the following two conditions:

1) The value of risk is stable (e.g. the growth of risk in the past 5000 searches is less than 0.1% of current risk value).

2) The total probability of simulated cascade paths exceeds a certain threshold (e.g. 0.97).

Usually the distribution of risks among all possible cascade paths is extremely non-uniform that risk is concentrated in a small portion of cascading outage paths. Observing the risk (2), since risk assessment is to add terms onto (2), the searching for states with larger risk terms in priority will achieve faster convergence of risk. To accelerate the convergence of the risk, a strategy of searching should be proposed which guides searching to the paths with major contribution to the risk. Next we will establish strategies that guide searching to such states.

## IV. TREE SEARCH STRATEGY USING RISK ESTIMATION INDEX

### A. Risk Estimation Index

Take the partial Markovian tree in Fig. 8 to study the searching strategy. Assume that searching has reached "*" state (labelled as $(i_{k_1} \cdots i_{k_{r-1}})$) and is about to select a next-level event $i_{k_r}$ (hollow nodes pointed by solid line arrows) to simulate. The strategy should let the increment of risk of the selected path be as large as possible, so the first task is to estimate the risks of all the subsequent states with acceptable computation complexity.

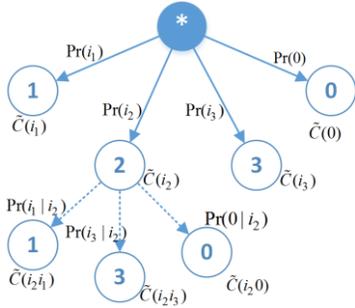

Fig. 8. Partial Markovian Tree when Search Starts

Here a risk estimation index $\rho_{i_{k_r}|i_{k_1} \cdots i_{k_{r-1}}}$ (simply denoted as $\rho_{i_{k_r}}$ since all studied subsequent states in this section have the same previous events $i_{k_1} \cdots i_{k_{r-1}}$ ) is established, and probabilities for searching are determined using $\rho_{i_{k_r}}$. The index consists of the following three parts:

#### 1) Risk of System Separation

If the outage of a branch causes the grid to separate, then the branch is called a cut branch of the grid. According to [37], cut branches can be identified with complexity of $O(|E|)$, where $|E|$ is the number of connected branches. Denote the admittance matrix of the grid as $Y$, then its Penrose-Moore pseudo-inverse uniquely exists, denoted as $Z$

$$Z = Y^+ \quad (7)$$

A branch $i_{k_r} = \{u, v\}$ is a cut branch if and only if

$$Y_{uv}^{-1} - 2Z_{uv} + Z_{uu} + Z_{vv} = 0 \quad (8)$$

Considering numerical errors, set a sufficiently small threshold $\varepsilon$ (e.g. $10^{-10}$ ), if

$$\left| Y_{uv}^{-1} - 2Z_{uv} + Z_{uu} + Z_{vv} \right| < \varepsilon \quad (9)$$

then the branch is identified as a cut branch. If $i_{k_r} = \{u, v\}$ is a cut branch and it fails, the separated two parts of the system will have unbalanced power $\pm F_{uv}$, which needs power balancing and generates cost. Therefore, the cost of system separation caused by cut branch $i_{k_r}$ outage is estimated as

$$\sigma_{i_{k_r}}^{\alpha} = \begin{cases} 2|F_{uv}|, \left| Y_{uv}^{-1} - 2Z_{uv} + Z_{uu} + Z_{vv} \right| < \varepsilon \\ 0, \left| Y_{uv}^{-1} - 2Z_{uv} + Z_{uu} + Z_{vv} \right| \ge \varepsilon \end{cases}, i_{k_r} = \{u, v\} \quad (10)$$

And the risk of system separation is estimated as

$$\rho_{i_{k_r}}^{\alpha} = \Pr(i_{k_1} \cdots i_{k_r}) \sigma_{i_{k_r}}^{\alpha} \quad (11)$$

#### 2) Risk of Overloading

After the outage of non-cut branches, the power flow will re-distribute throughout the system and may cause overloading on other elements, leading to costs generated by re-dispatch or emergency control actions. The influence of a branch outage on other branches can be quantified by the power transfer distribution factor (PTDF). The PTDF of a non-cut branch $\{u, v\}$ to any other branch $\{p, q\}$ [37] is

$$\delta_{uv}^{pq} = -\frac{Z_{up} + Z_{vq} - Z_{uq} - Z_{vp}}{1 + Y_{uv}(Z_{uu} + Z_{vv} - 2Z_{uv})} Y_{pq} \quad (12)$$

The power flow on $\{p, q\}$ after the outage of $\{u, v\}$ is

$$F_{pq}^{*uv} = F_{pq} + \delta_{uv}^{pq} F_{uv} \quad (13)$$

And the extent of overloading on branch $\{p, q\}$ is

$$\pi_{pq}^{*uv} = \max\{|F_{pq}^{*uv}| - F_{pq}^{\max}, 0\} \quad (14)$$

Define the overloading index of branch $\{u, v\}$ outage as

$$\sigma_{i_{k_r}}^{\beta} = \sum_{\{p, q\} \in E} \pi_{pq}^{*uv}, i_{k_r} = \{u, v\} \quad (15)$$

and define the estimation of overloading risk as

$$\rho_{i_{k_r}}^{\beta} = \Pr(i_{k_1} \cdots i_{k_r}) \sigma_{i_{k_r}}^{\beta} \quad (16)$$

Observing the denominator of (9), if $\{u, v\}$ is a cut branch, then the denominator is 0. So the PTDF of a cut branch has no definition. Therefore if (9) is satisfied, then $\rho_{i_{k_r}}^{\beta} = 0$.

For any studied state, about $|E|^2$ PTDF values and post-outage flows on branches are needed to calculate, so the estimation of overloading risk has the complexity of $O(|E|^2)$.

#### 3) Secondary Risk

Considering Fig. 8, when selecting the next-level states of "*" state, the risk of subsequent states of next-level states should also be accounted for. This risk is called a secondary risk in this paper. Since the secondary risk is hard to analytically quantify, a rough estimation is given in this paper.

First calculate the power flow $F_{pq}^{*uv}$ of all the connected lines $\{p, q\}$ after outage of branch $i_{k_r} = \{u, v\}$, and calculate the corresponding probabilities of outage $\Pr_{pq}^{*uv}$ during the next interval using (4)-(5). According to the overloading extent of $F_{pq}^{*uv}$, give an estimation of the cost $\tilde{C}_{pq}^{*uv}$ (it is difficult to accurately analyze, so in this paper $\tilde{C}_{pq}^{*uv}$ is set as 1% of system load), then the secondary risk of $i_{k_r} = \{u, v\}$ is

$$\rho_{i_{k_r}}^{\gamma} = \Pr(i_{k_1} \cdots i_{k_r}) \left( \sum \Pr_{pq}^{*uv} \tilde{C}_{pq}^{*uv} / \left| E_{i_{k_1} \cdots i_{k_r}} \right| \right) \quad (17)$$

where $E_{i_{k_1} \cdots i_{k_r}}$ is the set of connected branches at state $(i_{k_1} \cdots i_{k_r})$, and $\left| E_{i_{k_1} \cdots i_{k_r}} \right|$ is the number of connected branches.

If a next-level state has no outage, i.e. $i_{k_r} = 0$, then system separation risk and overloading risk are both considered as 0, but the secondary risk may be non-zero. In this case, if approximately regarding the system state at $i_{k_r} = 0$ the same as that of $i_{k_{r-1}}$, then the secondary events can be seen as shifting the next-level events of $i_{k_{r-1}}$ to $i_{k_r}$. If the probability of $i_{k_r} = 0$ is $\Pr_0$, then the corresponding secondary risk can be defined as

$$\rho_0^{\gamma} = \mu \frac{\Pr_0}{\left| E_{i_{k_1} \cdots i_{k_{r-1}}} \right|} \sum_{i_{k_r}} \rho_{i_{k_r}} \quad (18)$$

where $\mu \le 1$ is a discount factor considering that risk will be reduced by control schemes in the system.

For a studied state, the complete estimation of secondary risks needs about $|E|^2$ times of $\text{Pr}_{pq}^{*uv}$ and $\tilde{C}_{pq}^{*uv}$ calculation, so the complexity of secondary risk estimation is $O(|E|^2)$.

4) Establishing the risk estimation index

Till now, at arbitrary newly-searched state $(i_{k_1} \cdots i_{k_{r-1}})$, the risk estimation index of next-level branch $i_{k_r} = \{u, v\}$ outage is

$$\rho_{i_{k_r}} = \alpha \rho_{i_{k_r}}^{\alpha} + \beta \rho_{i_{k_r}}^{\beta} + \gamma \rho_{i_{k_r}}^{\gamma} \quad (19)$$

where $\alpha, \beta, \gamma$ are weights of risk terms. In this paper we select $\alpha = \beta = \gamma = 1$. The risk estimation index of $i_{k_r} = 0$ is

$$\rho_0 = \gamma \rho_0^{\gamma} \quad (20)$$

The above derivation of risk estimation index in this paper only considers branch outages, while the methodology of establishing risk estimation index can be used for the risk estimation of other kinds of events. Since in power flow model, the system state after bus/generator outage can be similarly derived using distribution factor, so the risk estimation index of bus and generator outages can be similarly established. However, the instability events have significantly different mechanism from the derivations above, therefore the severity estimation of instability events may be different from the above analysis. To limit computational complexity of risk estimation, stability indices [38-42] can be utilized to estimate the severity [43, 44] of instability event.

The overall complexity of risk estimation index is $O(|E|^2)$, while in the simulation of each level of cascading outages, the update of matrices is $O(|V|^2) \sim O(|V|^3)$, and the complexity of re-dispatch is bounded by $O(|E|^{3.5})$. Therefore, the calculation of risk estimation index is much lower than the simulation, so the calculation of risk estimation index does not notably affect the overall efficiency.

B. Forward-Backward Scheme of Markovian Tree Search

1) Forward searching using risk estimation index

As shown in Fig. 8, if a new state (labeled with asterisk) is reached on the Markovian tree, then all the subsequent states and paths are new. The risk estimation indices of next-level states are calculated and probabilities for selecting these states can be obtained using risk estimation indices. If the index is thought as accurate reflection of risks, then the optimal search strategy is to guide to the path with the highest risk estimation index value, which is a deterministic strategy

$$\text{Pr}_{i_{k_r}}^{calc} = \begin{cases} 1, i_{k_r} = \arg\max_i \{\rho_i\} \\ 0, otherwise \end{cases} \quad (21)$$

However, the risk estimation index may have error, so it is essential to have randomness in searching. Another strategy is random search with equal probability

$$\text{Pr}_{i_{k_r}}^{calc} = 1 / \left( \left| E_{i_{k_1} \cdots i_{k_{r-1}}} \right| + 1 \right) \quad (22)$$

The searching strategies of (21) and (22) represent two extremes: deterministic search vs. pure random search. To keep the merits of both approaches, the strategy can be selected in between. Introduce a parameter $\lambda \geq 0$ and set probability

$$\text{Pr}_{i_{k_r}}^{calc} = \frac{\left( \rho_{i_{k_r}} \right)^{\lambda}}{\sum\limits_{\xi \in E_{i_{k_1} \cdots i_{k_{r-1}}} \cup \{0\}} \left( \rho_{\xi} \right)^{\lambda}} \quad (23)$$

For $\lambda = 0$, equation (23) is equivalent to (22), and for $\lambda \to +\infty$, equation (23) is approximately (21).

During the simulation of cascading outages, the matrices $Y$, $Z$ need to be updated. If a set of branches $\{i_k\}$ are removed from the grid, then the admittance matrix can be updated with

$$Y' = Y - M_{\{i_k\}} Diag(y_{\{i_k\}}) M_{\{i_k\}}^T \quad (24)$$

where $M_{\{i_k\}}$ is a $|V| \times |\{i_k\}|$ matrix. Each of its column $M_{i_k}$ corresponding to a branch $i_k = \{u, v\}$ satisfies $M_{i_k, u} = 1$, $M_{i_k, v} = -1$ and all other entries are 0. $Diag(y_{\{i_k\}})$ is a matrix with branch admittances of $\{i_k\}$ 's as its diagonal elements. The update of matrix $Y$ (24) can be finished with a very small amount of calculation, with complexity of $O(|\{i_k\}|)$. The update of $Z$ is realized with

$$Z' = Z + ZM_{\{i_k\}} z_{\{i_k\}}^{-1} M_{\{i_k\}}^T Z \quad (25)$$

where

$$z_{\{i_k\}} = Diag(y_{\{i_k\}})^{-1} - M_{\{i_k\}}^T ZM_{\{i_k\}} \quad (26)$$

Since outages usually occur to very few branches at a time, $z_{\{i_k\}}$ is small and (25) has complexity of $O(|V|^2)$. However the inverse of (26) requires that $\{i_k\}$ is not a cut set. If $\{i_k\}$ is a cut set, then the update of $Z$ has to be realized with SVD of $Y$, which has complexity of $O(|V|^3)$ [37]. Theoretically, the SVD is computationally expensive, and an alternative that searches islands and simulate events on each island respectively consumes less computational resources. However, tests on RTS-96 system show that the instances requiring SVD only accounts for less than 20% of the matrix update computations, so the influence of SVD on overall efficiency is not very significant. Moreover, the actual efficiency may depend on realization. Take the realization on MATLAB as example, the SVD is a mature and efficient built-in function and supports further enhancement such as GPU acceleration, so the matrix with SVD update is also a practical and convenient option.

2) Backward updating risk estimation indices

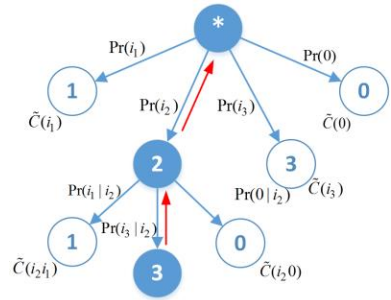

Fig. 9. Backward Update of risk estimation indices on Markovian tree

After states on cascading outage paths are visited and recorded, reaching these states again in the future will not contribute to the increment of the risk. Therefore, after a cascading outage path is found, the risk estimation indices should also be updated. On the contrary to the searching direction from the root to terminals of the Markovian tree, the updating of risk estimation indices should go backwards from terminals to the root. As shown in Fig. 9, assume that solid nodes are visited states and the node labelled as 3 in the bottom is the terminal of the path. Since searching to a visited terminal again will not make any contribution to the risk, then for a terminal state $i_{k_n}$ on a cascading outage path $i_k i_{k_2} \cdots i_{k_n}$, assign its risk estimation index a sufficiently small value $\rho_{i_{k_n}} = \varepsilon_R$ to avoid visiting it again.

For a non-terminal state $i_{k_r}$, since it has been visited and the risk term on the state $i_{k_r}$ itself will not contribute to the risk again, then its risk estimation index will only reflect risks of its subsequent states. Since risk estimation indices of all its next-level states $\rho_{i_{k_{r+1}} | i_{k_1} i_{k_2} \cdots i_{k_r}}$ must have been calculated or even updated, and the probabilities for searching are $\mathrm{Pr}_{i_{k_{r+1}}}^{calc}$ according to (23), then the risk estimation index of $i_{k_r}$ is

$$\rho_{i_{k_r}} = \sum_{i_{k_r}'} \rho_{i_{k_{r+1}} | i_{k_1} i_{k_2} \cdots i_{k_r}} \mathrm{Pr}_{i_{k_{r+1}}}^{calc} \qquad (27)$$

Equation (27) represents recursive backward updating of risk estimation indices. In the risk assessment on Markovian tree, first do forward search along a path, and then reversely update risk estimation indices using (27).

### C. Procedures of risk assessment with Markovian tree search

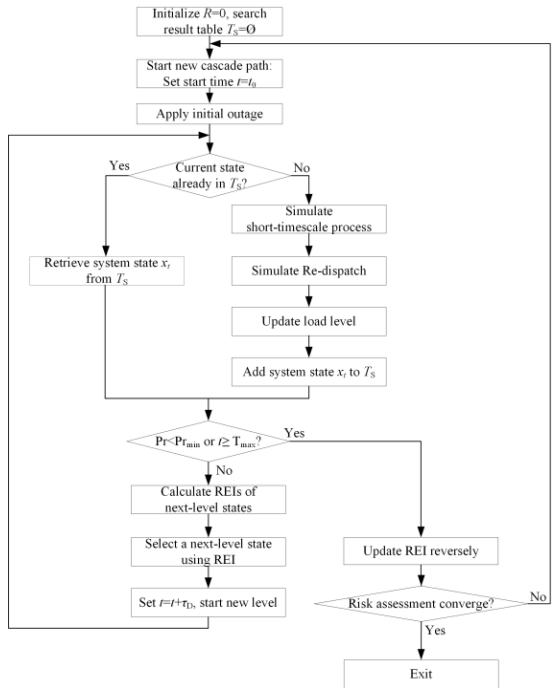

Fig. 10. Flowchart of risk assessment

Fig. 10 demonstrates the procedures of the proposed risk assessment method. To realize non-duplicated system state search, a search result table $T_S$ is established to index and store the states that have been searched. During risk assessment, if a state to be simulated is found in $T_S$, then the state is directly retrieved and the simulation of this level of events is avoided. If the state is not found in $T_S$, then this level of outage is simulated according to Section II.

## V. CASE STUDIES

### A. Illustrative 4-Node System

First verify the accuracy of the method in this paper with a small scale 4-node system. The system has 5 branches, 2 generator nodes and 2 load nodes, as shown in Fig. 11.

#### 1) Verification of Risk Assessment Performance

Set outage of branch 2-3 as the initial failure, and use the proposed method to assess the post-failure risk of the system. Set $T_{\max} = 60\,\mathrm{min}$ and $\tau_D = 15\,\mathrm{min}$, then the number of intervals is $K_D = \lceil T_{\max} / \tau_D \rceil = 4$, where $\lceil \cdot \rceil$ is "ceil" operation. With $N = 4$ and $K_D = 4$ we can get $N_T = 73$ according to (6), which is not a big number and we can calculate the theoretical value of the risk through the enumeration of all possible cascade paths as $R = 0.14208\,\mathrm{MW}$.

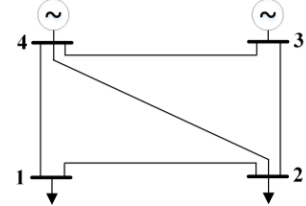

Fig. 11. Structure of 4-Node System

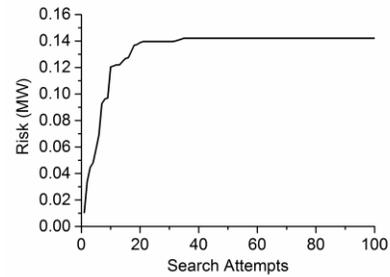

Fig. 12. Variation of Risk ($\lambda$=5)

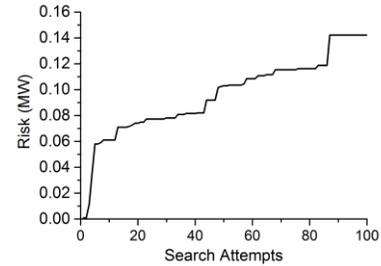

Fig. 13. Variation of Risk ($\lambda$=0.01)

Then use the method proposed in this paper. Set $\lambda = 5$ and $\lambda = 0.01$ respectively, and search on the Markovian tree for up to 100 paths, the variation of the risk as searching continues are shown in Figs. 12 and 13 separately. A search attempt means the search and simulation of an entire cascading outage path. In both cases, the risk reaches theoretical value $R = 0.14208\,\mathrm{MW}$, but the speeds of the risk convergence are distinct. The case with larger $\lambda$ achieves a faster convergence profile. In the case with $\lambda = 5$, the convergence of the risk only takes 35 search attempts, while the case of $\lambda = 0.01$ takes 87 attempts.

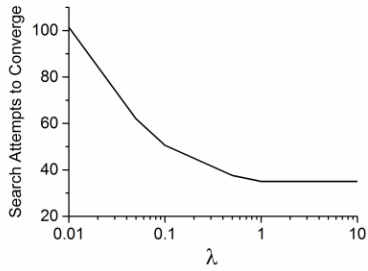

Fig. 14. Relationship between Required Attempts to Convergence and $\lambda$

Fig. 14 demonstrates the average search attempts to convergence under different values of $\lambda$. In this case, risk assessment is repeated 50 times under each $\lambda$ value. It is shown that with the increase of $\lambda$, the required number of search attempts decreases and finally stays at 35 times. To further assess the convergence of the risk with $N_S$ search attempts, construct the following convergence metric

$$\varphi_{N_S} = \sum_{j=1}^{N_S} j \cdot (R_{N_S} - R_j) \tag{28}$$

where $R_j$ is the risk after $j$ th search attempt and $R_{N_S}$ is the converging risk after $N_S$ search attempts. A smaller $\varphi_{N_S}$ metric means a better convergence profile.

TABLE I. CONVERGENCE METRIC UNDER DIFFERENT TRUST FACTORS

| $\lambda$ | Average Search Attempts | Average value of $\varphi_{Ns}$ | Standard deviation of $\varphi_{Ns}$ |
|---|---|---|---|
| **0.01** | 101.4 | 145.98 | 59.34 |
| **0.05** | 62 | 53.76 | 16.80 |
| **0.1** | 50.6 | 29.33 | 7.647 |
| **0.5** | 37.6 | 10.58 | 2.261 |
| **1** | 35 | 8.144 | 1.276 |
| **5** | 35 | 6.176 | 0.315 |
| **10** | 35 | 5.994 | 0.214 |
| **500** | 35 | 6.045 | 0.235 |
| **2000** | 35 | 5.834 | 0.145 |
| **10000** | 35 | 5.871 | 0.256 |

Table I and Fig. 15 demonstrate convergence metrics under different values of $\lambda$. It shows that a larger $\lambda$ gets a smaller $\varphi_{N_S}$ so as to achieve a better convergence profile. Moreover, by rearranging the sequence of all cascade paths, a theoretically optimal sequence with minimum $\varphi_{N_S}$ can be obtained for evaluating convergence of the risk assessment. In this case the theoretically minimum is $\varphi_{N_S}^* = 3.01$. Compared with the $\varphi_{N_S}$ under a nearly random search ($\lambda = 0.01$), the $\varphi_{N_S}$ of search in which risk estimation index plays more important role under $\lambda \geq 1$ is much closer to $\varphi_{N_S}^*$, and the deviation of $\varphi_{N_S}$ is much smaller, which means more stable performance.

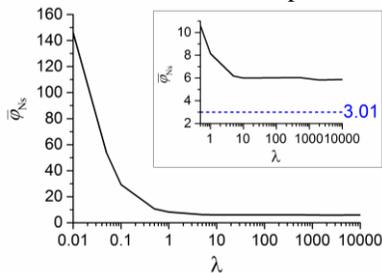

Fig. 15. Convergence metric under values of different $\lambda$

Analyzing the mechanism of Markovian tree search, a larger $\lambda$ tends to select cascading outage paths with higher risk estimation indices. The effectiveness of the proposed tree search scheme also verifies that the risk estimation indices can well reflect the distribution of actual risks. Since risk estimation index has a low computation complexity, the practicality of risk estimation index is verified.

However, the selection of $\lambda$ is not simply "the larger the better" since a too large $\lambda$ annihilates the randomness in path selection, and thus might miss some "hidden" risky states. Moreover, a too large $\lambda$ may cause overflow of floating point numbers in (23). Therefore in this case selecting $\lambda$ within 1~100 is suggested.

### 2) Influence of Mid-timescale Interval Length

The mid-timescale interval length $\tau_D$ is an important parameter in Markovian tree search. The determination of $\tau_D$ was discussed in [24], but since there are some difference between [24] and this paper in detail, the influence of $\tau_D$ should also be studied. Assign $\tau_D$ from 3min to 20min, and do risk assessment respectively.

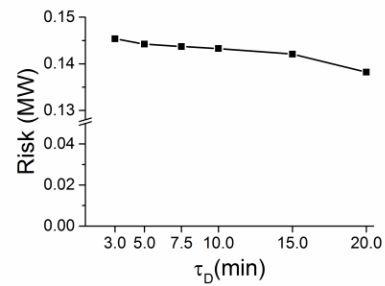

Fig. 16. Influence of different $\tau_D$ on risk assessment results.

From Fig. 16, although the value of $\tau_D$ are quite different, the assessed risks are fairly close, showing insensitivity of results to the $\tau_D$, which is reasonable since $\tau_D$ is a simulation parameter. As $\tau_D$ increases, risk slightly decreases because a larger $\tau_D$ allows fewer outages in the same period of time, so that some cases of multiple outages cannot be covered and the assessed risk is lower. But from the results in Fig. 16, the contribution of successive multiple outages to risk is rather limited since the probability of such a case is usually small. So a larger $\tau_D$ can satisfy the requirement of accuracy and has the advantage of better efficiency. If accuracy is preferred, then a smaller $\tau_D$ is desirable, while if efficiency is the priority, a larger $\tau_D$ is more suitable with satisfactory accuracy.

### 3) The Impact of Re-dispatch Delay

This risk assessment method can be utilized to study the impact of some system parameter on the cascading outage risk, and gives a clue on how to lower the risk.

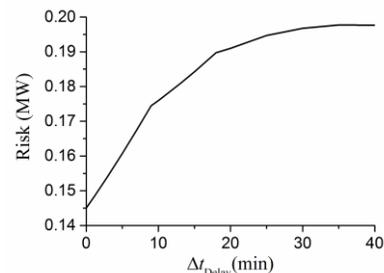

Fig. 17. The impact of re-dispatch delay on risk

The operation delay is an important performance factor related to system security but is seldom studied in existing research on cascading outages. Here, we change $\Delta t_{Delay}$ to study the impact of delay on risk as shown in Fig. 17. It is shown that the risk rises as $\Delta t_{Delay}$ increases, and the impact of delay is more significant when $\Delta t_{Delay}$ is small. Because the re-dispatch is usually activated in 5-10 minutes [27] after events, the risk can be more effectively reduced if delay is shortened. In practice the decrease in delay usually requires a control system upgrade, and the results in Fig. 17 can be helpful in the cost/effect analysis for system upgrade.

### B. RTS-96 Test System

#### 1) Performance of risk estimation index

In this part the method is tested on a larger RTS-96 3-area system [12] having 73 buses, 120 branches, 33 generator nodes and 51 load nodes. Select outages of branches 22, 23 and 24 as initial outages and assess risk. In this case, set $\tau_D = 15$ min, $T_{max} = 150$ min, delay of re-dispatch $\Delta t_{Delay} = 30$ min.

First we test the accuracy of the proposed risk estimation index in estimating the risk of subsequent cascading outage paths. Since the risk estimation index is proposed to facilitate the selection of next-level cascading outage path, we mainly compare the risk estimation indices of the states having the same previous state. For example, we study the risk estimation indices and the subsequent risk of the states on level 1. The subsequent risk of level-1 state $i_{k_1}$ is

$$R(i_{k_1}) = \Pr(i_{k_1})C(i_{k_1}) + \Pr(i_{k_1})\sum_{k_2}\Pr(i_{k_2} \mid i_{k_1})C(i_{k_1}i_{k_2})$$
$$+ \Pr(i_{k_1})\sum_{k_2}\Pr(i_{k_2} \mid i_{k_1})\sum_{k_3}\Pr(i_{k_3} \mid i_{k_1}i_{k_2})C(i_{k_1}i_{k_2}i_{k_3}) + \cdots \quad (29)$$

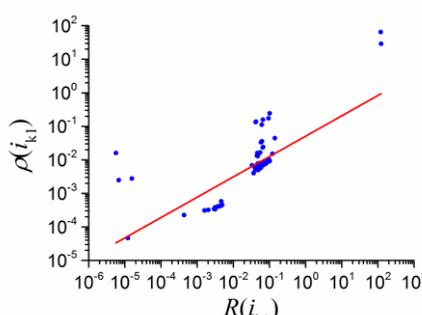

Fig. 18. Correlation between risk and risk estimation index

Fig. 18 shows the relationship between subsequent risk of level-1 state $R(i_{k_1})$ and risk estimation index $\rho_{i_{k_1}}$ in a log-log plot. The plot contains 118 scattered data points. The results visually demonstrate positive relationship between $R(i_{k_1})$ and $\rho_{i_{k_1}}$. Linear regression of these points show the approximated quantified relationship as

$$\log_{10} \rho_{i_{k_1}} = 0.6063 \cdot \log_{10} R(i_{k_1}) - 1.2964$$

and the Pearson's correlation coefficient is 0.712, indicating strong linear positive correlation between $R(i_{k_1})$ and $\rho_{i_{k_1}}$. The risks and risk estimation indices on other levels of Markovian tree generally show similar strong positive correlation, which verifies that the risk estimation index can effectively guide the tree search to cascading outage paths with higher risks.

#### 2) Efficiency test of risk assessment

According to (6), the number of possible cascading paths is about $3.5491 \times 10^{20}$, so it is unrealistic to enumerate them all and calculate theoretical risk value. Here use a relatively large number of search attempts $N_S$ and regard the risk at $N_S$ attempts $R_{Ns}$ as the theoretical risk $R$. The test program is developed and tested in MATLAB on a workstation with 2.6 GHz processor and 32GB RAM.

Set $N_S = 300000$ and get risk $R_{Ns} = 252.76$MW through risk assessment. From Fig. 19, the risk rises sharply at the beginning and approaches $R_{Ns}$ quickly in the first several thousands of search attempts, and then its rising speed becomes much slower. As Table II shows, after 19 search attempts the risk has reached $0.5R_{Ns}$, and then after 2709 attempts the index reaches $0.9R_{Ns}$, with computation time less than 10 min. But reaching $0.99R_{Ns}$ takes a much larger amount of computation, consuming several hours. From the perspective of application, it is of practically required to apply risk assessment with limited computation time. In this case, no more than 5000 attempts and 1000 seconds of computation time can account for more than 90% of the cascading risk.

Fig. 20 shows the covered probability of simulated cascade paths along the process of risk assessment. At the beginning, the coverage of probability rises sharply over 0.9, and after the first 2000 search attempts nearly 0.97 has been covered. Therefore, most probable cascade paths have been simulated and assessed, and since the risk estimation index can effectively guide computation to cascade paths with major risks, the rest of the paths are expected to have minor contribution to the risk.

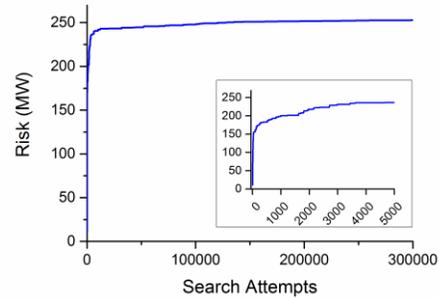

Fig. 19. Risk on RTS-96 test system

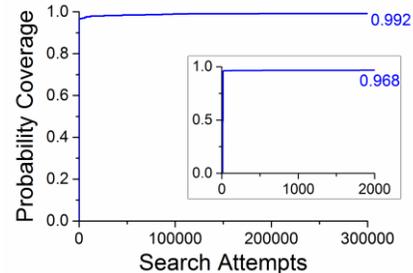

Fig. 20. Coverage of probability in risk assessment

TABLE II. EFFICIENCY TEST OF RTS-96 SYSTEM

| >=$R_{Ns}$% | 50% | 90% | 95% | 99% | 99.9% |
|---|---|---|---|---|---|
| Search Attempts | 19 | 2709 | 6259 | 129134 | 259856 |
| Time (s) | 3.87 | 552 | 1277 | 26360 | 53044 |
| Prob. Coverage | 0.961 | 0.969 | 0.973 | 0.991 | 0.992 |

With a high-performance computer and software-level optimization, the computation efficiency of this method is expected to meet the need for online applications. Moreover, using risk assessment results, all the simulated cascading outage paths and risks can be analyzed and measures for lowering risk can be established [45], which constitutes our future work.

As for the memory usage, in this case where 300000 cascading outages are simulated, about 1.8 million states are stored in the memory, and each state variable occupies 672bytes, so the total memory usage for recording states is about 1GB, which is easily satisfied on ordinary PCs. For larger systems, the requirement for memory space can also be satisfied on workstations, servers or other high-performance platforms. Therefore, even though the memory usage of the Markovian tree search method is higher than some existing methods, the requirement for memory space is generally affordable for practical use.

## VI. CONCLUSIONS

In this paper, a risk assessment method of multi-timescale cascading outages based on Markovian tree search is proposed. The method first equivalently reformulates our previous work of quasi-dynamic cascading outage simulation as a Markovian tree model, thus the method maintains the advantage of reasonable modeling and simulation for multi-timescale cascading outages. The Markovian tree model bridges the reasonable modeling and simulation with the efficient risk assessment. Then the methodology of risk assessment by non-duplicated cascading outage path searching on Markovian tree is proposed, which enhances efficiency by avoiding duplicated searches on cascade paths and effectively exploiting computation resources.

To accelerate the convergence of the risk, this paper proposed a risk estimation index that estimates the risks of next-level cascading outage paths with low computational efficiency, and a "forward search – backward update" scheme for risk assessment based on the risk estimation index is established. The strategy can effectively and efficiently guide the search to paths with major contributions to the risk, further enhancing efficiency of risk assessment.

The method is first tested on an illustrative 4-bus system to verify the accuracy and effectiveness of risk assessment by a comparison with theoretical results. The selection of search strategy is also tested and analyzed from the perspective of the balancing between the deterministic selection of the largest risk estimation index and random searching, and the effectiveness of risk estimation index is verified. The risk assessment method is also tested on the RTS-96 system, showing that the method is able to effectively search out riskiest cascade paths and states accounting for more than 90% risk in less than 10 minutes, indicating that the computation speed has potential to meet the requirements for operational risk assessment.

## ACKNOWLEDGEMENT

The authors would like to thank Prof. Janusz Bialek at the Skoltech Center of Energy Systems, the Skolkovo Institute of Science and Technology for hosting Rui Yao as a visiting scholar and providing many useful suggestions.